\documentclass[a4paper,11pt]{article} 
\usepackage{graphicx}
\usepackage{psfrag}
\usepackage{amsmath}
\usepackage{fleqn}
\usepackage{pslatex}
\usepackage[colorlinks,bookmarks,dvips]{hyperref}

\newcommand{\sigmaEff}{\sigma_{\mbox{\scriptsize eff}}}

\newcommand{\calT}{{\cal T}}
\newcommand{\EMA} {\mbox{EMA}}
\newcommand{\MA} {\mbox{MA}}

\newcommand{\dTref}{\Delta T_{\mbox{\scriptsize ref}}}
\newcommand{\dt}{\delta t}
\newcommand{\dtr}{\delta t_r}
\newcommand{\dts}{\delta t_\sigma}
\newcommand{\dtd}{\delta t_\Delta}
\newcommand{\sigmaDot}{\dot{\sigma}}

\begin{document} 
\begin{center}
{\LARGE\bf
       Heterogeneous volatility cascade \\ in financial markets
}

\vspace*{6ex}
{\Large Gilles Zumbach$\mbox{}^{a,}$\footnote{\parbox[t]{0.9\textwidth}{
    e-mail: gilles@olsen.ch\hspace{2em}  phone: +41-1/386 48 24\hspace{2em} Fax: +41-1/422 22 82}}
        and Paul Lynch$\mbox{}^{a,b}$
}

$\mbox{}^{a}$~{\footnotesize
    Olsen \& Associates,
    Research Institute for Applied Economics\\
    Seefeldstrasse 233, 8008 Z\"urich, Switzerland.
    }

$\mbox{}^{b}$~{\footnotesize
    University of Manchester Institute of Science and Technology,
    Electrical \& Electronic Engineering\\
    PO Box 88,
    Manchester, M60 1QD,
    England}

\vspace*{4ex}
February 22, 2001

\vspace*{10ex}

{\Large\bf Abstract}
\end{center}
Using high frequency data, 
we have studied empirically the change of volatility, also called 
volatility derivative, for various time horizons.
In particular, the correlation between the volatility derivative and the 
volatility realized in the next time period is a measure of the 
response function of the market participants.
This correlation shows explicitly the heterogeneous structure of the market according to 
the characteristic time horizons of the differents agents.
It reveals a volatility cascade from long to short time horizons,
with a structure different from the one observed in turbulence.
Moreover, we have developed a new ARCH-type model which incorporates 
the different groups of agents, with their characteristic memory.
This model reproduces well the empirical response function, 
and allows us to quantify the importance of each group.

\vspace*{8ex}
Turbulence.
\newpage

\section{Introduction}
Financial markets are very interesting self-organized structures.
On a given market, say for example the foreign exchange (FX) market for USD/CHF, 
a large number of agents are present.
These agents differ by their geographic locations, individual preferences, 
economic expectations, information sets, present market positions, 
educations, risk aversions or professional constraints. 
Yet, at a given time point, the market agree on one thing: a price.
An interesting categorization of the market participants 
can be found in their characteristic time frame: intra-day speculators and market makers, 
daily traders, portfolio managers rebalancing their positions every week,
or pension funds and central banks that are active at a scale of months 
using fundamental economics measures.
Although this categorization makes sense, until now it has gone unobserved.
Essentially, the only endogenous information available about a financial market is the 
resulting price $p$ as a function of time $t$, 
and the curve $p(t)$ looks like a random walk.

A related topic is the efficient market hypothesis (EMH) \cite{EFF.1970-01-01,MJe.1978-01-02}.
This hypothesis can be formulated in many different ways, with various strengths.
For example, a (semi) strong formulation can be 
"given all the publicly available information, the price process is a martingale",
and a weak formulation could be 
"there are no dependencies in past price changes that a
technician could use to predict future changes".
There exists a huge body of literature on this topic, with many empirical 
tests of particular formulations of the hypothesis.
This hypothesis is rooted in the rationality of the market participants:
humans are rational and behave in their best interests.
Within a strong formulation of the EMH, given (all) the information at time $t$,
each market participant should behave in the same rational way,
and the market should incorporate ``instantaneously'' 
every new information to reach a new equilibrium price.
This implies that the market participants behave as one group,
a picture quite different from the time characterization given above.
On the other hand, research on the microstructure of the FX market
conducted by questionnaires survey of dealers \cite{YHL.1998-01-01} indicates
a heterogeneous set of time horizons, 
and the practical importance of technical analysis.

Another piece of evidence related to the market composition is 
the recent analogy with fully developed turbulence \cite{SHG.1996-01-01}. 
These authors have compared the probability density function (pdf) of the 
return (i.e. price changes) for a set of time horizons $\delta t$ with the 
pdf of velocity differences in a fluid for a set of position differences.
The striking agreement of the pdf's leads to the conclusion that
the vorticity cascade responsible for turbulence should 
have a counterpart in financial processes.
Therefore, an information cascade must be present in financial market, 
from long time horizons up to intra-day traders.

The basic question underlying the above points is the homogeneous or 
heterogeneous composition of the financial market, 
as well as the possible different agent's time responses and mutual interactions. 
Beside some indirect evidences and arguments, this is essentially an open question.
An indirect evidence for the market structure has already been obtained 
in \cite{UAM.1997-01-01} through a log-likelihood estimate for the HARCH model
(with an induced market structure in agreement with the results below).
What is missing is a statistical estimate, derived from the the price process $p(t)$,
that is able to display the underlying structure of the market participants. 
By studying the volatility time derivative, 
we have found such a quantity in the correlation between the 
change of volatility and the realized volatility, 
thus providing a visual proof of a heterogeneous market structure.
The paper proceeds as follows:
the next section will introduce the definitions we are using.
Then, we will present empirical results, followed by a model for the 
volatility process that incorporates the structure of the market.

\section{The volatility and its time derivative}
Our data processing starts with tick by tick quotes obtained from Reuter.
These quotes arrive at random time $t$ and contain a bid and ask price.
From the quotes, we compute the logarithmic middle price 
$x(t) = 0.5\;(\ln(\mbox{bid}(t)) + \ln(\mbox{ask}(t)))$.
These raw mid-prices are then smoothed by a very short term moving 
average with a range of 3 minutes in order to eliminate the tick by tick noise.

High frequency data contain very strong intra-day and intra-week seasonalities, 
namely a predictable repetitive pattern due to the daily and weekly cycles of human activity.
These seasonalities are filtered out by doing the computations
in the proper business time scale. 
The key idea is similar to the usual business time scale used when working with daily data, 
namely to simply omit the week-ends and major holidays from the computations.
The time scale we are using is an improvement along this idea, 
namely to expand periods of high activity and to contract periods of low activity (week-end, night).
The seasonal activity pattern of high frequency data is measured on a moving sample,
and the dynamic time scale is constructed by integrating the activity.
The time scale is normalized such that, on average, a time interval of $\delta t$ 
in physical time scale is equal to $\delta t$ in business time.
The basic ideas are presented in \cite{MMD.1992-01-01}, 
and the dynamic algorithm we are using is explained in detail in \cite{WAB.2000-01-31}, 
including the discounting of holidays and the treatment of daylight saving time.
Let us emphasize that the proper discounting of the 
seasonalities is a mandatory preliminary step in order to obtain the results presented in this paper.

Given the dynamic time scale, we compute a regular 
time series $x(i)$ of smoothed prices where the sampling is 
done every $\dt =$ 10 minutes on the dynamic time scale.
From this regular time series, the historical volatility is computed with
\begin{eqnarray}
    r[\dtr](i) & = & \frac{x(i) - x(i - 1)}{\sqrt{\dtr/\dTref}}  \label{eq:r}\\
    \sigma_h^2[\dts, \dtr](i) & = & \frac{1}{n} \sum_{i - p + 1 \leq j \leq i} r^2[\dtr](j)
\end{eqnarray}
with $\dtr = \dt$,
$\dts = p\, \dt$, $n = \sum_{i - p + 1 \leq j \leq i}$.
The denominator in eq.~\ref{eq:r} ``annualizes'' the return, 
namely discounts the random walk scaling such that the expectation $E[~ r^2[\dtr] ~]$ is essentially 
independent of $\dtr$, with a typical value of 10\% for FX rates.
The reference time interval $\dTref$ is taken to be one year.
The volatility derivative $\sigmaDot$ is computed using a smooth difference kernel
according to \cite{GOZ.1998-11-01} applied on 
the historical volatility
\begin{equation}
    \sigmaDot[\dtd, \dts, \dtr] = \Delta\left[\dtd; \sigma_h[\dts, \dtr]\right].
\end{equation}
The operator $\Delta$ essentially computes a finite difference 
$\Delta[\dtd; z](t) \simeq z(t) - z(t - \dtd)$,
but using a convolution with a smooth kernel instead of the difference of pointwise values.
Let us emphasize that the notation $\sigmaDot$ for the volatility derivative
is indeed referring to a finite difference at a time scale $\dtd$.
This smooth derivative at a finite time scale is the appropriate
notion of derivative for a random process, 
namely it measures the mean change at a time scale $\dtd$.
For this article, in order to reduce the dimension of the parameter space,
we restrict ourselves to $\dtd = \dts$ and $\dtr = \dt$,
but other choices lead to similar results.
With this choice for the parameters, we can use the shorter notation 
$\sigma_h[\dts] = \sigma_h[\dts, \dt]$ and $\sigmaDot[\dtd] = \sigmaDot[\dtd, \dtd, \dt]$.
In the finance literature, the analysis of the volatility derivative is new,
as researchers have focused until now on the return and volatility.
The volatility derivative $\sigmaDot$ is a particularly interesting quantity as it measures 
dynamical aspects of the volatility evolution.
A full analysis of the statistical properties of the 
volatility derivative is presented in \cite{GOZ.2001-02-30}.

The historical volatility and volatility derivative at time $t$ 
are computed using information in the past up to time $t$.
The realized volatility corresponds to the ``next'' volatility after $t$,
namely is computed from prices in the future of $t$.
Using a forward time translation operator $\calT[\dt; x](t) = x(t+\dt)$, 
the realized volatility is
\begin{equation}
   \sigma_r[\dts] = \calT[\dts; \sigma_h[\dts]]
\end{equation}
and we have again restricted ourselves to $\dtr = \dt$.

We have also explored other definitions for the volatility and the derivative.
The volatility can be defined as an aggregated volatility with 
$r[\dtr](i) = (x(i) - x(i - k))/\sqrt{\dtr/\dTref}$
with $\dtr = k\; \dt$. 
The derivative can be taken with a logarithm, namely
$\sigmaDot[\dtd, \dts, \dtr] = \Delta\left[\dtd; \ln(\sigma_h[\dts, \dtr]) \right]$.
With all these definitions, very similar results are obtained, 
both for the empirical and simulated correlation.

\section{The market response function}
We have computed the usual linear correlation between the volatility derivative 
$\sigmaDot[\dtd]$ and the realized volatility $\sigma_r[\dts]$,
for time intervals ranging from 4 hours to 42 days.
This correlation measures the response function of 
the market to a change of volatility, 
similar to the phenomenological susceptibility
introduced in electro-magnetism with matter for example.
As the market participants react to changes of volatility at a given time scale, 
they may change their positions and induce volatilities in the next time period.
The correlation $\rho(\dtd, \dts) = \rho[\sigmaDot[\dtd], \sigma_r[\dts]]$
measures this response function.

\begin{figure}
  \centering
  \psfrag{vol der}{\begin{large} $\dtd$ \end{large}}
  \psfrag{real vol}{\begin{large} $\dts$ \end{large}}
  \psfrag{corr}{\begin{large} $\rho$ \end{large}}
  \includegraphics[width=0.8\textwidth]{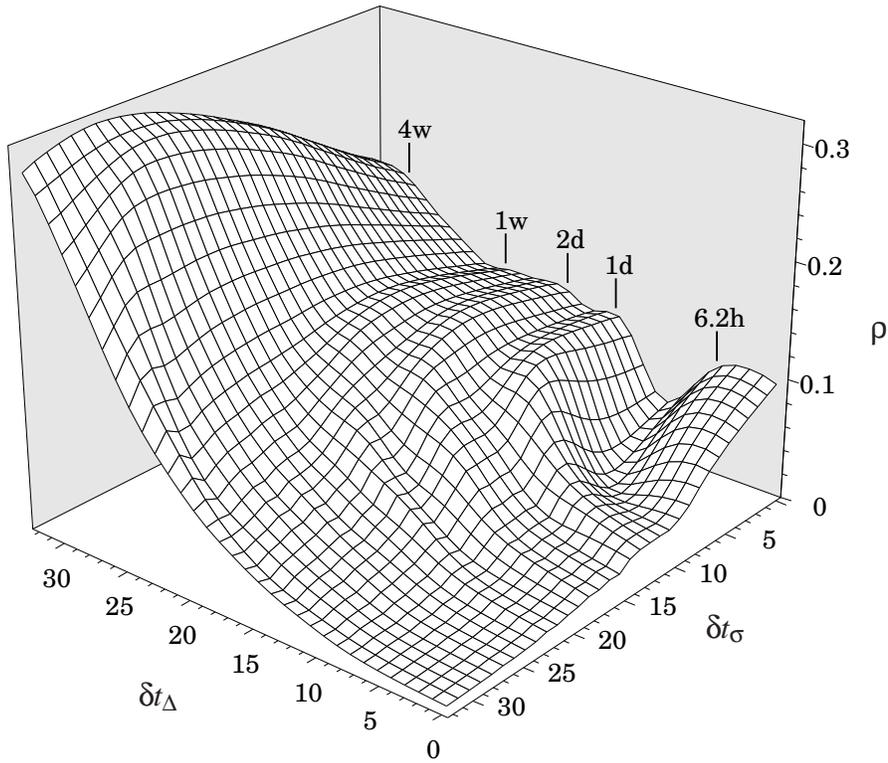}
  \caption{\sf
    The empirical correlation between volatility derivative and realized volatility 
    $\rho[\sigmaDot[\dtd], \sigma_r[\dts]]$ for the foreign exchange USD/CHF.
    The business time intervals corresponding to the axis label $k$ are given by $\dt = 2^{k/4}\;4$ hour, 
    and span 4 hours (k=0) to 6 weeks (k=32).
    The data sample used to compute the correlation ranges from 01.01.1990 to 01.11.2000, 
    and the dynamic time scale and volatility are initialized using data from 1.7.1988 to 31.12.1989.
  }
  \label{fig:USDCHF}
\end{figure}

The computed correlation for 10 years of USD/CHF is displayed in Fig.~\ref{fig:USDCHF},
and clearly shows different groups of market participants. 
At short time scales, intra-day traders quickly react to short term change of volatility.
However, short term volatility changes 
do not induce a response from traders with longer time horizons. 
Changes in the volatility at the daily time scale 
trigger the response of both intra-day and daily traders, 
but not of the weekly and longer horizons market participants.
Notice the gap between 8 hours to 1 day, corresponding to the absence 
of traders working inside this time frame. 
Then, slow change in the volatility induces a response of all market participants 
working at shorter time scales, while the maximal response is at a similar time horizon.
The correlation is always positive, which means that
the market participants react to an increase of volatility by changing positions
(and therefore they increase the realized volatility),
but they are not likely to react to a decrease of volatility.
Overall, the pattern that emerges is similar to a volatility cascade from 
low frequency to high frequency \cite{AAR.1997-01-01}, but with changes in volatility 
triggering the response of all shorter time horizons.
This is different from the picture in turbulence where the vorticity 
at a given time scale is related to vorticity only at nearby time scales.

For other currency pairs or for stock indexes \cite{GOZ.2001-02-30}, a similar structure emerges.
However, there are quantitative differences, the most important one being a smaller
cluster corresponding to intra-day traders for stock indexes. 
This can be understood from the higher cost of trading stocks 
(brokerage costs and larger bid-ask spread), 
making it less profitable to trade intra-day.
Finally, the correlation between historical and realized volatility can be computed
$\rho[\sigma_h[\dts], \sigma_r[\dts']]$.
This correlation is dominated by the heteroskedasticity of the financial market,
namely by the long memory (or the clustering) of the volatility.
A finer structure due to the market components lies on top
of the overall heteroskedasticity, but the structure of the market
does not appear clearly in the volatility-volatility correlation.
It is only the response induced by changes in volatility 
that reveals the components of the market.

\section{Modeling the market components}
In order to fully understand the above picture, it is interesting 
to compare the empirical correlation with the one obtained with Monte Carlo simulations 
of theoretical processes.
The simplest theoretical model is an i.i.d. random walk.
This model has a zero correlation $\rho(\dtd, \dts) = 0$.
A better benchmark, widely used in finance, is the GARCH(1,1) model \cite{TBO.1986-01-01}. 
A Monte Carlo simulation for this model shows a positive correlation, 
with one weakly defined maximum located around the correlation time of the process.
This is clearly inadequate to model a market with sharply defined components.
We have developed a new model, called Market-Component-ARCH($n$) or MC-ARCH($n$) model,
in order to reproduce a market with $n$ components.
Structurally, this model draws from GARCH(1,1),  
the long memory model presented in \cite{GOZ.2000-02-14}, 
and the HARCH model \cite{UAM.1997-01-01,MMD.1998-01-01}.  
It is built using iterated exponential 
moving averages that induce
a sharp cut-off for the memory of each component \cite{GOZ.1998-11-01}. 
The MC-ARCH model equations are as follows:
\begin{eqnarray}
    x(t+\dt) & = & x(t) + r(t+\dt)        \label{eq:x}\\
    r(t+\dt) & = & \sigmaEff(t+\dt)\; \epsilon(t+\dt)  \label{eq:rp}  \\
    \sigmaEff^2(t+\dt) & = & w_\infty\; \sigma^2 + \sum_{k = 1}^n w_k \sigma_k^2(t) 
              \label{eq:sigmaEff} \\
    \sigma_k^2(t) & = & \MA[\tau_k, m; r^2](t)               \label{eq:sigmaK}
\end{eqnarray}
with the constraint on the coefficients
\begin{equation}
    w_\infty + \sum_{k = 1}^n w_k = 1~.                           \label{eq:sumW}
\end{equation}
The time interval $\dt$ fixes the time scale at which the process is defined.
Eq.~\ref{eq:x} says that the logarithm of the price $x = \ln(p)$ 
follows a random walk with price increment $r$.
From eq.~\ref{eq:rp}, at each time step, the return $r$ is the product of a 
magnitude $\sigmaEff$ and a random variable $\epsilon$.
The random variable $\epsilon(t)$ is independent and identically distributed (i.i.d.), 
with the conditions $E[\epsilon(t)] = 0$ and $E[\epsilon^2(t)] = 1$.
For the simulations, we have taken a Student-t distribution with $\nu = 5$ degree of freedom,
a number consistent with an estimate obtained through a maximum likelihood optimization.
The magnitude $\sigmaEff(t+\dt)$ can be seen as a forecast for 
the effective volatility of the market at $t+\dt$. 
This forecast is build using the information available at $t$ (eq.~\ref{eq:sigmaEff}).
The constant $\sigma$, with the constraint~\ref{eq:sumW}, fixes the mean volatility of the process, 
namely $E[r^2(t)] = E[\sigmaEff^2(t)] = \sigma^2$.
The mean volatility is the volatility measured at an infinite time scale,
and therefore its amplitude is denoted by $w_\infty$.
The volatility $\sigma_k$ is measured by a moving average (MA) 
at the time scale $\tau_k$ of the squared returns (eq.~\ref{eq:sigmaK}). 
Essentially, this term models the perceived current price volatility for a market 
participant with a memory of depth $\tau_k$ in the past.
The volatility $\sigma_k$ contributes with a weight $w_k$ 
to the effective volatility (eq.~\ref{eq:sigmaEff}).
The model parameters are $\tau_k$, $w_k$ and $\sigma$.
The MA operator \cite{GOZ.1998-11-01} for the time range 
$\tau$ acting on the time series $z$ is defined as
\begin{eqnarray}
    \MA[\tau, m; z](t) & = & \frac{1}{m} \sum_{j=1}^m \EMA_j(t)        \label{eq:MA} \\
    \EMA_1(t)  & = & \mu\EMA_1(t-\dt) + (1 - \mu)z(t)       \label{eq:sigma0} \\
    \EMA_j(t)  & = & \mu\EMA_j(t-\dt) + (1 - \mu)\EMA_{j-1}(t)  \label{eq:sigmaJ}\\
    \mu        & = & \exp\left(-\;\delta t(m+1)/\tau\right)                  \label{eq:mu}
\end{eqnarray}
with the shorthand notation $\EMA_j = \EMA[\tau, j; z]$.
The coefficients $\mu$ (eq.~\ref{eq:mu}) is computed from the time horizon $\tau$,
so that the memory length of the MA operator is $\tau$. 
Technically, the memory length is twice the range of the kernel 
of the corresponding MA operator \cite{GOZ.1998-11-01},
a measure appropriate for rectangular like kernels.
The MA operator is computed through a sum of iterated exponential moving average (EMA) 
(eq.~\ref{eq:sigma0} and \ref{eq:sigmaJ}).
The coefficient $m$ controls the shape of the decay for the memory, from exponential ($m=1$)
to rectangular ($m\rightarrow\infty$). 
Practically, $m=32$ is already close to a rectangular memory.

The structure of the MC-ARCH model is similar to the HARCH model 
\cite{UAM.1997-01-01,MMD.1998-01-01} in that both include 
several volatilities measured on a set of time horizons. 
Yet, the HARCH model was developed mainly to include the asymmetry in the response 
function of the volatility measured with returns at different time horizons $r[\dtr]$, 
whereas we find this effect to be quantitatively unimportant.
On the other side, it is important to have the proper time horizons for each market component,
as well as the correct memory decay for the volatility measure, 
features not contained in the HARCH model.

\begin{figure}
  \centering
  \psfrag{volDer}{\begin{large} $\dtd$ \end{large}}
  \psfrag{realVol}{\begin{large} $\dts$ \end{large}}
  \psfrag{corr}{\begin{large} $\rho$ \end{large}}
  \includegraphics[width=0.8\textwidth]{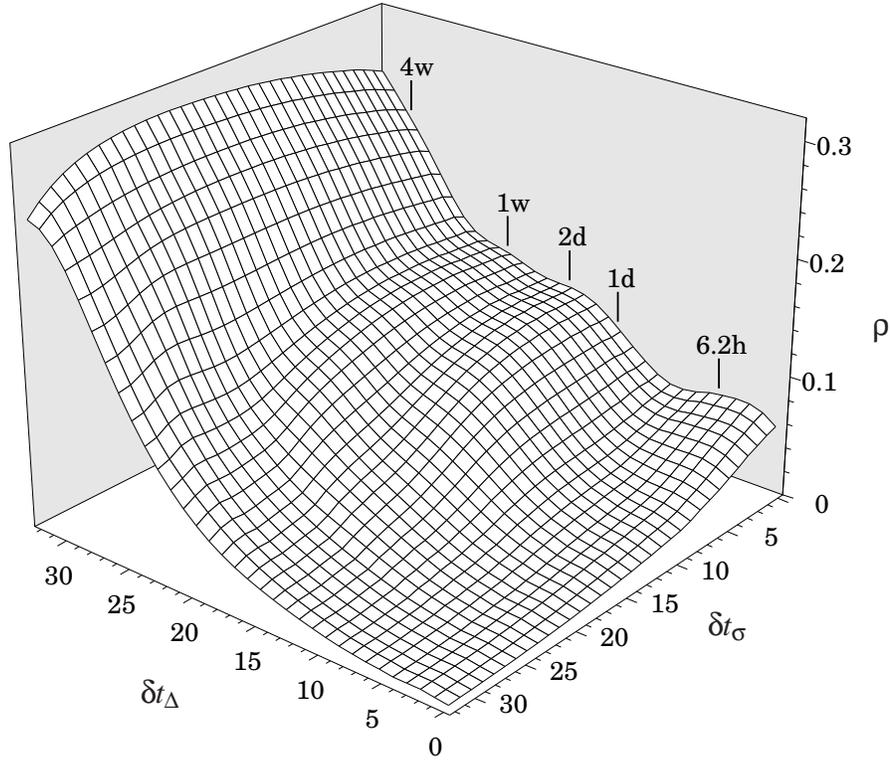}
  \caption{\sf
    The correlation between volatility derivative and realized volatility 
    $\rho[\sigmaDot[\dtd], \sigma_r[\dts]]$ for the MC-ARCH(5) model, computed by Monte Carlo simulations.
    The business time interval corresponding to the axis label $k$ are given by $\dt = 2^{k/4}\;4$ hour, 
    and span 4 hours (k=0) to 6 weeks (k=32).
    The labels on the backdrop correspond to the characteristic 
    time of each component of the MC-ARCH model, expressed in physical time 
    (for time intervals shorter than a week, a factor 5/7 is used to map business time 
    intervals to physical time intervals in order to discount for the week-end).
    The length of the simulation is $10^6$ steps, corresponding to 19 years.
  }
  \label{fig:MC-ARCHCorr}
\end{figure}

The parameters of the model have been optimized 
by simulations so as 
to reproduce the empirical figure for the 
correlation $\rho[\sigmaDot[\dtd], \sigma_r[\dts]]$.
Good results are obtained by taking 5 components with characteristic 
times (measured in business days) 
$\tau_k =$ 0.18 (intra-day), 1.4 (1 day), 2.8 (2 days), 7 (1 weeks) and 28 (4 weeks).
The correlation obtained by simulation is given in Fig.~\ref{fig:MC-ARCHCorr}
and the agreement with the empirical correlation is excellent.
The coefficient $m$ controlling the shape of the volatility kernel
has to be taken high enough, for the figure $m=64$.
For $m=1$, the shape of simulated correlation is too ``soft'', 
as it does not show the empirical abrupt drop to zero 
or the separation between intra-day and daily traders.
This large value for $m$ can be interpreted as an abrupt decay of the memory of 
the corresponding market component, 
namely the actors forget quickly the past beyond their characteristic time scale. 
The coefficients for each component $w_k$ are respectively  0.39, 0.20, 0.18, 0.12, 0.11, and $w_\infty =$ 0.00025.
If we interpret these coefficients as measuring the 
``financial weight'' of the respective component, 
we see that the largest 
fraction of the FX market is carried by short term  dealers (intra-day, daily).
Quantitatively, the actors with a characteristic time horizon 
up to two days account for 76\% of the market.

The MC-ARCH model does not contain an explicit term with $\sigmaDot$.
A ``pure'' volatility model with the correct market structure is enough 
to reproduce the main feature of the empirical 
correlation $\rho[\sigmaDot[\dtd], \sigma_r[\dts]]$.
Yet, we are not able to reproduce the sharp ``valley'' between intra-day and daily horizon, 
nor the decay of the correlation below the diagonal in the realized volatility direction.
Possibly, a $\sigmaDot$ term can be added in the MC-ARCH model 
in order to better match the empirical correlation, hence opening a whole new space of models.

\section{Conclusion}
The correlation between the change of volatility and realized volatility 
gives a picture of the market components and their responses.
The pattern that emerges is that a change of volatility at a given time scale
triggers a response, and therefore volatility, at all shorter time scales.
The response function is clustered around values corresponding 
to well defined group of market participants, like intra-day dealers,
portfolio managers or pension funds.
This picture is a bit different from fully developed turbulence 
where the vorticity cascade relates nearby scales:
the turbulence at a given scale is feed by the scale right 
above and feeds the scale right below.
Moreover, the cascade is homogeneous.
In financial markets, a change at a given time scale feeds all 
the shorter time horizons, and the structure is heterogeneous.

The MC-ARCH($n$) model presented here incorporates the relevant market structure, 
and has the same response function as observed in empirical data.
It allows us to quantify the importance of each group, 
and shows that the agents quickly forget the past 
beyond their characteristic time horizons.
Finally, to our amazement, from the apparently random walk of the price,
we are able to extract only by statistical means a 
clear picture of the market heterogeneity.

\bibliographystyle{unsrt}

\end{document}